\documentclass[3p,times]{elsarticle}
\usepackage{amssymb}
\usepackage{amsxtra}
\usepackage{amsmath}
\usepackage{amsfonts}
\usepackage{CJK}
\usepackage[titletoc]{appendix}
\usepackage{graphicx}
\usepackage{hyperref}
\usepackage{cleveref}
\numberwithin{equation}{section}

\newtheorem{thm}{Theorem}[section]

\newtheorem{prop}[thm]{Proposition}

\newtheorem{ass}[thm]{Assumption}

\newtheorem{de}[thm]{Definition}

\newcommand{\eqa}{\begin{eqnarray}}
\newcommand{\eeqa}{\end{eqnarray}}
\newcommand{\beq}{\begin{equation}}
\newcommand{\eeq}{\end{equation}}
\newcommand{\nn}{\nonumber}

\allowdisplaybreaks
\begin{document}
\begin{frontmatter}
\title{On the Riemann-Hilbert problem for the Chen-Lee-Liu derivative nonlinear Schr\"{o}dinger equation}
\author[BB]{Beibei Hu\corref{cor1}}
\author[BB]{Ling Zhang\corref{cor1}}
%\author[TC]{Tiecheng Xia}
\author[ZN]{Ning Zhang}
%\ead{This research is partially supported by NSF of China (Grant numbers 11372170).}

\cortext[cor1]{Corresponding authors at: School of Mathematics and Finance, Chuzhou University, Anhui 239000, China. \\
 Email addresses: hu\_chzu@shu.edu.cn(B.-B. Hu), originzhang@126.com(L. Zhang)}

\address[BB]{School of Mathematics and Finance, Chuzhou University, Anhui, 239000, China}
%\address[TC]{Department of Mathematics, Shanghai University, Shanghai 200444, China}
\address[ZN]{Department of Basical Courses, Shandong University of Science and Technology, Taian 271019, China}
\pagestyle{plain}
\setcounter{page}{1}
\begin{abstract}
In this work, we investigated a combined Chen-Lee-Liu derivative nonlinear Schr\"{o}dinger equation(called CLL-NLS equation by Kundu) on the half-line by unified transformation approach. We gives spectral analysis of the Lax pair for CLL-NLS equation, and establish a matrix Riemann-Hilbert problem, so as to reconstruct the solution $r(z,t)$ of the CLL-NLS equation by solving. Furthermore, the spectral functions are not independent, but enjoy by a compatibility condition, which is the so-called global relation.
\end{abstract}
\end{frontmatter}

\section{Introduction}
As we all know, the nonlinear Schr\"{o}dinger(NLS) equation and the derivative NLS(DNLS) equation are all important equations in the field of mathematical physics. On the one hand, the NLS equation\cite{BDJ1967}
\eqa ir_{t}+r_{zz}\pm|r|^2r=0, \label{1.0}\eeqa
can be used to describe the movement of microscopic particles, and is also one of the basic equations of quantum mechanics. In quantum mechanics, the solution of particle problems is usually converted into the solution of the stationary NLS equation problem.
On the other hand, there are three types of DNLS equation, the first is the famous Kaup-Newell equation\cite{KN1978}, also known as DNLS-I equation
\eqa r_{t}+r_{zz}\pm i(|r|^2r)_z=0, \label{1.1a}\eeqa
the second is Chen-Lee-Liu equation\cite{CLL1979}, also known as DNLS-II equation
\eqa ir_{t}+r_{zz}-i|r|^2r_x=0, \label{1.1b}\eeqa
and the last is Gerdjikov-Ivanov equation\cite{GI1983}, also known as DNLS-III equation
\eqa ir_{t}+r_{zz}-i|r|^2\bar{r}_x+\frac{1}{2}|r|^4r=0. \label{1.1c}\eeqa
The DNLS type equation has important applications in nonlinear optics and plasma physics. It can be used not only to describe picosecond pulse in single-mode nonlinear fiber, but also to control the evolution of small amplitude Alfv\'{e}n waves or large amplitude magnetohydrodynamic(MHD) waves in plasmas.

With the in-depth study of soliton theory, a unified transformation(UT) approach to solve the initial-boundary value problems(IBVPs) of integrable model is proposed by Fokas\cite{Fokas1997}, also known as the Fokas approach. Since the UT approach have been proposed, the IBVPs of many nonlinear integrable models have been extensively studied\cite{FAS2004,Lenells2011,Lenells2012,Lenells2013,Monvel2013,Xu2013,Xu2014,Xia2017,Tian2017,AI2018,ZYS2019,Yan2019,CSY2019,HL2020,HB2020}.
In particular, Fokas et al. analyzed the IBVPs of the NLS equation\eqref{1.0} on the half-line\cite{Fokas2005}, Monvel et al. gives the IBVPs on a finite interval case\cite{Boutet2006}. Lenells discussed the IBVPs of the Kaup-Newell(DNLS-I) equation\eqref{1.1a} on the half-line \cite{Lenells2008}, Xu and Fan gives the IBVPs on a finite interval case\cite{XJ2014}. Zhang et al. studied the IBVPs of the Chen-Lee-Liu(DNLS-II) equation\eqref{1.1b} on the half-line\cite{ZN}. Zhu et al. investigated the IBVPs of the Gerdjikov-Ivanov(DNLS-III) equation\eqref{1.1c} on the half-line\cite{ZQZ2017}, and then gives the IBVPs on a finite interval\cite{ZQZ2018}.

In 1984, Kundu\cite{KA1984} proposed a combined Chen-Lee-Liu derivative nonlinear Schr\"{o}dinger equation (called CLL-NLS equation by Kundu)
\eqa ir_{t}+r_{zz}+|r|^2r-|r|^2r_{z}=0. \label{1.2}\eeqa
which is a completely integrable model. In fact, the CLL-NLS equation\eqref{1.2} can be derived from the modified NLS equation(also known as the Dysthe equation) which ignores the mean flow term in hydrodynamics\cite{DKB1979}. In 2014, Chan et al. give the rogue wave of the CLL-NLS equation\eqref{1.2} based on the Hirota bilinear transformation, and pointed out that the CLL-NLS model can be explain the experimental phenomena in nonlinear optical fibers and water wave flumes\cite{CHN2014}.  Especially, Zhang et al. obtain higher-order solutions of the CLL-NLS equation\eqref{1.2} by using the Darboux transformation method\cite{ZYS2015}, includes non-vanishing boundary solitons, breathers and rogue wave solutions.
In this paper, let $(z,t)\in\Gamma=\{(z,t)|0<z<\infty,0<t<T\}$, we aim to investigate the IBVPs of the CLL-NLS equation\eqref{1.2} via UT approach.

The paper is organized as follows. Section 2, we will gives spectral analysis of the Lax pair for \eqref{1.2}. Section 3, some key functions $u(\varsigma), v(\varsigma), U(\varsigma), V(\varsigma)$ are further analyzed. Section 4, the Riemann-Hilbert problem is proposed. Section 5 are some conclusions and discussions.

\section{The spectral analysis}

The Lax pair for CLL-NLS equation \eqref{1.2} as follows\cite{ZYS2015,CPA1987,LX2013}
\begin{subequations}
\begin{align}
&\Psi_{z}=M(z,t,\varsigma)\Psi=(i\sigma \varsigma^2-\frac{1}{2}i\sigma+R\varsigma+\frac{1}{2}iR^2\sigma)\Psi,\label{2.1a}\\
&\Psi_{t}=N(z,t,\varsigma)\Psi=[-2i(\varsigma^2-\frac{1}{2})^2\sigma-2R\varsigma^3-iR^2\sigma \varsigma^2+(R+i\sigma R_z-\frac{1}{2}R^3)\varsigma\nn\\
&\qquad\qquad\qquad\quad-\frac{1}{8}iR^4\sigma+\frac{1}{4}(RR_z-R_zR)]\Psi,\label{2.1b}
\end{align}
\end{subequations}
with
\beq
\Psi(z,t,\varsigma)=\left(\begin{array}{c}
\Psi_1(z,t,\varsigma)\\
\Psi_2(z,t,\varsigma)
\end{array}\right),
\sigma=\left(\begin{array}{cc}
1 & 0\\
0 & -1
\end{array}\right),
R=\left(\begin{array}{cc}
0 & r\\
-\bar{r} & 0
\end{array}\right). \label{2.2}\eeq

\subsection{The exact 1-form}

The Lax pair Eq.\eqref{2.1a}-\eqref{2.1b} rewrite as
\begin{subequations}
\begin{align}
&\Psi_{z}-i(\varsigma^2-\frac{1}{2})\sigma \Psi=A\Psi,\label{2.3a}\\
&\Psi_{t}+2i(\varsigma^2-\frac{1}{2})^2\sigma \Psi=B\Psi,\label{2.3b}
\end{align}
\end{subequations}
where the complex number $\varsigma$ is a associated spectral parameter and
$$A=R\varsigma+\frac{1}{2}iR^2\sigma,\,
B=-2R\varsigma^3-iR^2\sigma \varsigma^2+(R+i\sigma R_z-\frac{1}{2}R^3)\varsigma-\frac{1}{8}iR^4\sigma+\frac{1}{4}(RR_z-R_zR).$$

Let's do the first function transformation $\Phi(z,t,\varsigma)$ as follows
\eqa \Phi(z,t,\varsigma)=\Psi(z,t,\varsigma)e^{[-i(\varsigma^2-\frac{1}{2})z+2i(\varsigma^2-\frac{1}{2})^2t]\sigma},0<x<\infty, 0<t<T. \label{2.4}\eeqa
Hence, Eq.\eqref{2.3a}-\eqref{2.3b} equal to
\begin{subequations}
\begin{align}
&\Phi_z-i(\varsigma^2-\frac{1}{2})[\sigma,\Phi]=A\Phi,\label{2.6a}\\
&\Phi_t+2i(\varsigma^2-\frac{1}{2})^2[\sigma,\Phi]=B\Phi,\label{2.6b}
\end{align}
\end{subequations}
where $[\sigma,\Phi]=\sigma\Phi-\Phi\sigma$. It is not difficult to find that the above equations satisfies a full differential form
\eqa d(e^{[-i(\varsigma^2-\frac{1}{2})z+2i(\varsigma^2-\frac{1}{2})^2t]\bar{\sigma}}\Phi(z,t,\varsigma))
=e^{[-i(\varsigma^2-\frac{1}{2})z+2i(\varsigma^2-\frac{1}{2})^2t]\bar{\sigma}}(Adz+Bdt)\Phi(z,t,\varsigma),\label{2.7}\eeqa
where $\hat{\sigma}$ represents a matrix operator (see \cite{HB2020}).

One assume that the solution of Eq.\eqref{2.6a}-\eqref{2.6b} enjoy the following asymptotic expansion form
We suppose that the following asymptotic expansion
\beq \Phi(z,t,\varsigma)=Q_0+\frac{Q_1}{\varsigma}+\frac{Q_2}{\varsigma^2}+\frac{Q_3}{\varsigma^3}+O(\frac{1}{\varsigma^4}), \varsigma\rightarrow\infty.\label{2.9}\eeq
one substituting Eq.\eqref{2.9} into Eq.\eqref{2.6a} and Eq.\eqref{2.6b}, respectively, and comparing the coefficient for $\varsigma$ gives rise to
\begin{subequations}
\begin{align}
& Q_{0z}=\frac{1}{4}i|r|^2\sigma Q_0,\label{2.11a}\\
& Q_{0t}=(\frac{1}{8}|r|^4+\frac{1}{4}(\bar{r}r_z-r\bar{r}_z))\sigma Q_0.\label{2.11b}
\end{align}
\end{subequations}
Owing to Eq.\eqref{2.1a}-\eqref{2.1b} admits the conservation laws as follows
 $$(i|r|^2)_t=(\frac{1}{2}i|r|^4+(\bar{r}r_z-r\bar{r}_z))_z.$$
then, one can define
\eqa Q_0(z,t)=e^{i\int_{(\infty,0)}^{(z,t)}\Theta(z,t)\sigma}, \label{2.14}\eeqa
where $\Theta$ is the exact closed 1-form defined by
\eqa \Theta(z,t)=\Theta_1dz+\Theta_2dt=\frac{1}{4}|r|^2dx+
(\frac{1}{8}|r|^4-\frac{1}{4}i(\bar{r}r_z-r\bar{r}_z)) dt.\label{2.14a}\eeqa

As a result of Eq.\eqref{2.14} is independent of the integration path and $\Theta$ is independent of $\varsigma$, one can take the second function transformation $H(z,t,\varsigma)$
\eqa \Phi(z,t,\varsigma)=e^{i\int_{(0,0)}^{(z,t)}\Theta\hat{\sigma}}H(z,t,\varsigma)Q_0(z,t). \label{2.15}\eeqa
thus, Eq.\eqref{2.7} become to
\eqa d(e^{[-i(\varsigma^2-\frac{1}{2})z+2i(\varsigma^2-\frac{1}{2})^2t]\hat{\sigma}}H(z,t,\varsigma))=J(z,t,\varsigma), \label{2.16}\eeqa
where
\begin{subequations}
\begin{align}
&J(z,t,\varsigma)=e^{[-i(\varsigma^2-\frac{1}{2})z
+2i(\varsigma^2-\frac{1}{2})^2t]\hat{\sigma}}F(z,t,\varsigma)H(z,t,\varsigma),\label{2.17a}\\
&F(z,t,\varsigma)=A_1(z,t,\varsigma)dz+B_1(z,t,\varsigma)dt
=e^{-i\int_{(0,0)}^{(z,t)}\Theta\hat{\sigma}}(Adz+Bdt-i\Theta\sigma).\label{2.17b}
\end{align}
\end{subequations}
where $A(z,t,\varsigma), B(z,t,\varsigma)$ and $\Theta$ given by
\begin{subequations}
\begin{align}
&A_1(z,t,\varsigma)=\left(\begin{array}{cc}
-\frac{3i}{4}|r|^2 &  \varsigma re^{2i\int_{(0,0)}^{(z,t)}\Theta} \\
-\varsigma\bar{r} qe^{-2i\int_{(0,0)}^{(z,t)}\Theta} & \frac{3i}{4}|r|^2
\end{array}\right),\label{2.17a}\\
&B_1(z,t,\varsigma)=\left(\begin{array}{cc}
-\frac{i}{4}|r|^4+\frac{5}{4}(\bar{r}r_z-\bar{r}_zr) & (-2\varsigma^3r+i\varsigma^2|r|^2+\varsigma(r_z+r-|r|^2r))e^{2i\int_{(0,0)}^{(z,t)}\Theta}\\
(2\varsigma^3\bar{r}-i\varsigma^2|r|^2+\varsigma(\bar{r}_z-\bar{r}+|r|^2\bar{r}))e^{-2i\int_{(0,0)}^{(z,t)}\Theta}
 & \frac{i}{4}|r|^4-\frac{5}{4}(\bar{r}r_z-\bar{r}_zr)
\end{array}\right).\label{2.17b}
\end{align}
\end{subequations}
Then Eq.\eqref{2.16} equal to
\begin{subequations}
\begin{align}
&H_x-i(\varsigma^2-\frac{1}{2})[\sigma,H]=A_1H,\label{2.18a}\\
&H_t+2i(\varsigma^2-\frac{1}{2})^2[\sigma,H]=B_1H.\label{2.18b}
\end{align}
\end{subequations}

\subsection{ The analytic and bounded eigenfunctions $\{H_j(z,t,\varsigma)\}_1^3$}

Set that $r(z,t)\in \mathcal{S},((z,t)\in\Gamma)$, one enjoy three eigenfunctions $\{H_j(z,t,\varsigma)\}_1^3$ of Eq.\eqref{2.18a}-\eqref{2.18b} defined by following integral equation
\eqa
H_j(z,t,\varsigma)=\mathrm{I}+\int_{(z_j,t_j)}^{(z,t)}
e^{[i(\varsigma^2-\frac{1}{2})z-2i(\varsigma^2-\frac{1}{2})^2t]\hat{\sigma}}J(\zeta,\tau,\varsigma),
\label{2.19}\eeqa
where $(z_1,t_1)=(0,0), (z_2,t_2)=(0,T), (z_3,t_3)=(\infty,t)$. Since Eq.\eqref{2.19} has nothing to do with the integral path, one choose the integration path shown in Figure 1, then, one get
\begin{subequations}
\begin{align}
&H_1(z,t,\varsigma)=\mathrm{I}+\int_{0}^{z}e^{i(\varsigma^2-\frac{1}{2})(z-\zeta)\hat{\sigma}}(A_1H_1)(\zeta,t,\varsigma)d\zeta\nn\\
&\qquad\qquad\qquad
+e^{i(\varsigma^2-\frac{1}{2})z\hat{\sigma}}\int_{0}^{t}e^{-2i(\varsigma^2-\frac{1}{2})^2 (t-\tau)\hat{\sigma}}(B_1H_1)(0,\tau,\varsigma)d\tau,\label{2.19a}\\
&H_2(z,t,\varsigma)=\mathrm{I}+\int_{0}^{z}e^{i(\varsigma^2-\frac{1}{2})(z-\zeta)\hat{\sigma}}(A_1H_2)(\zeta,t,k)d\zeta\nn\\
&\qquad\qquad\qquad
-e^{-i(\varsigma^2-\frac{1}{2})z\hat{\sigma}}\int_{t}^{T}e^{-2i(\varsigma^2-\frac{1}{2})^2 (t-\tau)\hat{\sigma}_3}(B_1H_2)(0,\tau,\varsigma)d\tau,\label{2.19b}\\
&H_3(z,t,\varsigma)=\mathrm{I}-\int_{z}^{\infty}e^{i(\varsigma^2-\frac{1}{2})(z-\zeta)\hat{\sigma}3}(A_1H_3)(\zeta,t,\varsigma)d\zeta.\label{2.19c}
\end{align}
\end{subequations}

\begin{figure}
\centering
\includegraphics[width=4.0in,height=1.1in]{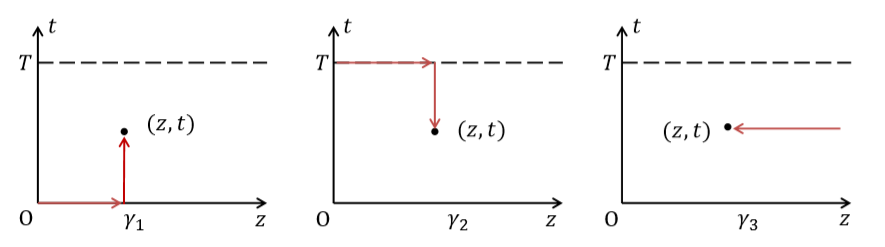}
\caption{The smooth curve $\gamma_1,\gamma_2,\gamma_3$ in the $(z,t)$-plan}
\label{fig:graph}
\end{figure}
Besides, the following equation holds on the $(x,t)$ plan
\begin{subequations}
\begin{align}
&\gamma_1=(z_1,t_1)\rightarrow(z,t): z-\zeta\geq 0, t-\tau\geq 0,\label{2.20a}\\
&\gamma_2=(z_2,t_2)\rightarrow(z,t): z-\zeta\geq 0, t-\tau\leq 0,\label{2.20b}\\
&\gamma_3=(z_3,t_3)\rightarrow(z,t): z-\zeta\leq 0.\label{2.20c}
\end{align}
\end{subequations}
In order to obtain the bounded analytical region of $\{H_j(z,t,\varsigma)\}_1^3$ on the complex $\varsigma$-plane, one use curve
$$\Omega(\varsigma)=\{\varsigma\in \mathbb{C}|\mathrm{Re}(\phi(\varsigma))\mathrm{Re}(\psi(\varsigma))=0,
\phi(\varsigma)=i(\varsigma^2-\frac{1}{2}),\psi(\varsigma)=2i(\varsigma^2-\frac{1}{2})^2\},$$
to divide the complex $\varsigma$-plane as shown in Figure 2.
Thus, we get
\eqa\begin{array}{l}
 D_1=\{\varsigma\in \mathbb{C}|\mathrm{Re}\phi(\varsigma)>0\,\,and \,\,\mathrm{Re}\psi(\varsigma)>0\},\\
 D_2=\{\varsigma\in \mathbb{C}|\mathrm{Re}\phi(\varsigma)>0\,\,and \,\,\mathrm{Re}\psi(\varsigma)<0\},\\
 D_3=\{\varsigma\in \mathbb{C}|\mathrm{Re}\phi(\varsigma)<0\,\,and \,\,\mathrm{Re}\psi(\varsigma)<0\},\\
 D_4=\{\varsigma\in \mathbb{C}|\mathrm{Re}\phi(\varsigma)<0\,\,and \,\,\mathrm{Re}\psi(\varsigma)>0\}.
\end{array}\label{2.21}\eeqa
Furthermore, the bounded analysis area of $\{H_j(z,t,\varsigma)\}_1^3$ as follows
\begin{subequations}
\begin{align}
& H_1(z,t,\varsigma):\,( \phi_{-}\cap\psi_{+},\phi_{+}\cap\psi_{-} )=:(D_4,D_2),\label{2.22a}\\
& H_2(z,t,\varsigma):\,( \phi_{-}\cap\psi_{-},\phi_{+}\cap\psi_{+} )=:(D_3,D_1),\label{2.22b}\\
& H_3(z,t,\varsigma):\,( \phi_{+},\,\phi_{-} )=:(D_1\cup D_2,D_3\cup D_4),\label{2.22c}
\end{align}
\end{subequations}
where $\phi_{+}=:\mathrm{Re}\phi(\varsigma)>0$, $\phi_{-}=:\mathrm{Re}\phi(\varsigma)<0$, $\psi_{+}=:\mathrm{Re}\psi(\varsigma)>0$, $\psi_{-}=:\mathrm{Re}\psi(\varsigma)<0$.

\begin{figure}
  \centering
  \includegraphics[width=2.0in,height=1.8in]{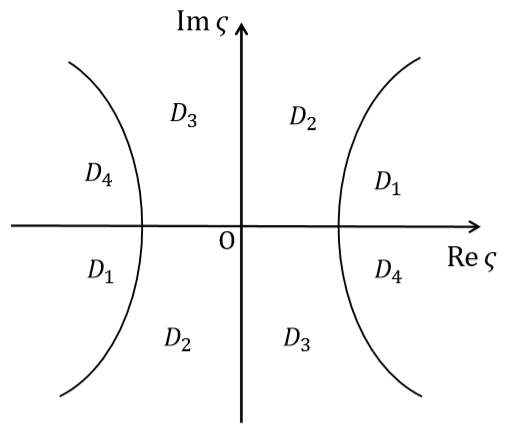}\\
  \caption{The areas $D_i,i=1,\ldots,4$ division on the complex $\varsigma$-plane}\label{fig:graph}
\end{figure}

To construct the Riemann-Hilbert problem of CLL-NLS equation \eqref{1.2}, one also need to define two important special functions $w(\varsigma)$ and $W(\varsigma)$ by:
\begin{subequations}
\begin{align}
&H_3(z,t,\varsigma)=H_1(z,t,\varsigma)e^{[i(\varsigma^2-\frac{1}{2})z-2i(\varsigma^2-\frac{1}{2})^2t]\hat\sigma}w(\varsigma),\label{2.24a}\\
&H_2(z,t,\varsigma)=H_1(z,t,\varsigma)e^{[i(\varsigma^2-\frac{1}{2})z-2i(\varsigma^2-\frac{1}{2})^2t])\hat\sigma}W(\varsigma).\label{2.24b}
\end{align}
\end{subequations}
Upon evaluation at $(z,t)=(0,0)$ and $(z,t)=(0,T)$, respectively, Eq.\eqref{2.24a} and Eq.\eqref{2.24b} yields
\eqa W^{-1}(\varsigma)=e^{-2i(\varsigma^2-\frac{1}{2})^2T\hat\sigma}H_1(0,T,\varsigma), w(\varsigma)=H_3(0,0,\varsigma),\label{2.25}\eeqa
According to \eqref{2.24a}-\eqref{2.24b} and Eq.\eqref{2.25}, one have
\eqa H_2(z,t,\varsigma)=H_3(z,t,\varsigma)e^{[i(\varsigma^2-\frac{1}{2})z-2i(\varsigma^2-\frac{1}{2})^2t]\hat\sigma}
(w(\varsigma))^{-1}W(\varsigma), \label{2.26}\eeqa
Especially, if one remember that $[H_j]_{k}(z,t,\varsigma)$ denotes $k$-columns of $H_j(z,t,\varsigma)$, one obtain $H_1(z,t,\varsigma), H_2(z,t,\varsigma)$ at $z=0$
\begin{subequations}
\begin{align}
&H_1(0,t,\varsigma)=([H_1]_1^{D_1\cup D_4}(0,t,\varsigma),[H_1]_2^{D_2\cup D_3}(0,t,\varsigma)) \nn\\
&\qquad\qquad\,=\mathrm{I}+\int_{0}^{t}e^{-2i(\varsigma^2-\frac{1}{2})^2 (t-\tau)\hat\sigma}(B_1H_1)(0,\tau,\varsigma)d\tau,\label{2.28a}\\
&H_2(0,t,\varsigma)=([H_2]_1^{D_2\cup D_3}(0,t,\varsigma),[H_2]_2^{D_1\cup D_4}(0,t,\varsigma))\nn\\
&\qquad\qquad\,=\mathrm{I}-\int_{t}^{T}e^{-2i(\varsigma^2-\frac{1}{2})^2(t-\tau)\hat\sigma}
(B_1H_2)(0,\tau,\varsigma)d\tau,\label{2.28b}
\end{align}
\end{subequations}
and $H_1(z,t,\varsigma),H_3(z,t,\varsigma)$ at $t=0$.
\begin{subequations}
\begin{align}
&H_1(z,0,\varsigma)=([H_1]_1^{D_3\cup D_4}(z,0,\varsigma),[H_1]_2^{D_1\cup D_2}(z,0,\varsigma))\nn\\
&\qquad\qquad\,=\mathrm{I}+\int_{0}^{z}e^{i(\varsigma^2-\frac{1}{2})(z-\zeta)\hat\sigma}(A_1H_1)(\zeta,0,k)d\zeta,\label{2.28c}\\
&H_3(z,0,\varsigma)=([H_{3}]_1^{D_1\cup D_2}(z,0,\varsigma),[H_{3}]_2^{D_3\cup D_4}(z,0,\varsigma))\nn\\
&\qquad\qquad\,=\mathrm{I}-\int_{z}^{\infty}e^{i(\varsigma^2-\frac{1}{2})(z-\zeta)\hat\sigma}(A_1H_3)(\zeta,0,\varsigma)d\zeta,\label{2.28d}
\end{align}
\end{subequations}
where $H_{j}^{D_i}$ represents the bounded analytic region of $\{H_j\}_1^3$ is $\varsigma\in D_i, i=1,2,3,4$.

Set that $r_0(z)=r(z,t=0)$, $s_0(t)=r(z=0,t)$, $s_1(t)=r_z(z=0,t)$ are initial data and boundary datas of $r(z,t)$ and $r_z(z,t)$, then, one yields
\begin{subequations}
\begin{align}
& A_1(z,0,\varsigma)=\left(\begin{array}{cc}
-\frac{3i}{4}|r_0|^2 &  \varsigma r_0e^{\frac{i}{2}\int_{0}^{z}|r_0|^2dz} \\
\varsigma\bar{r}_0e^{-\frac{i}{2}\int_{0}^{z}|r_0|^2dz} & \frac{3i}{4}|r_0|^2
\end{array}\right),\label{2.29a}\\
&B_1(0,t,\varsigma)=\left(\begin{array}{cc}
-\frac{i}{4}|s_0|^4+\frac{5}{4}(\bar{s}_0s_1-\bar{s}_1s_0) & B_{12}^{(1)}(0,t,\varsigma) \\
B_{21}^{(1)}(0,t,\varsigma) & \frac{i}{4}|s_0|^4-\frac{5}{4}(\bar{s}_0s_1-\bar{s}_1s_0)
\end{array}\right).\label{2.29b}
\end{align}
\end{subequations}
with
\eqa &&
B_{12}^{(1)}(0,t,\varsigma)=(-2\varsigma^3s_0+i\varsigma^2|s_0|^2+\varsigma(s_1+s_0-|s_0|^2s_0))
e^{\int_{0}^{t}(\frac{i}{4}|s_0|^4+\frac{1}{2}(\bar{s}_0s_1-s_0\bar{s}_1))dt},\nn\\&&
B_{21}^{(1)}(0,t,\varsigma)=(2\varsigma^3\bar{s}_0-i\varsigma^2|s_0|^2+\varsigma(\bar{s}_1-\bar{s}_0+|s_0|^2\bar{s}_1))
e^{-\int_{0}^{t}(\frac{i}{4}|s_0|^4+\frac{1}{2}(\bar{s}_0s_1-s_0\bar{s}_1))dt}
\nn\eeqa

\subsection{ The other properties of the eigenfunctions}
\begin{prop}
The functions $H_j(z,t,\varsigma)=([H_j]_{1}(z,t,\varsigma),[H_j]_{2}(z,t,\varsigma)), j=1,2,3, $ possess the following properties
\begin{itemize}
\item $\mathrm{det}H_j(z,t,\varsigma)=1, j=1,2,3$,
\item $[H_1]_{1}(z,t,\varsigma)$ is analytic for $\varsigma\in D_4$, and continues to $\bar{D}_4, [H_1]_{2}(z,t,\varsigma)$ is analytic for $\varsigma\in D_2$, and continues to $\bar{D}_2$.
\item $[H_2]_{1}(z,t,\varsigma)$ is analytic for $\varsigma\in D_3$, and continues to $\bar{D}_3, [H_2]_{2}(z,t,\varsigma)$ is analytic for $\varsigma\in D_1$, and continues to $\bar{D}_1$.
\item $[H_3]_{1}(z,t,\varsigma)$ is analytic for $\varsigma\in D_1\cup D_2$, and continues to $\bar{D}_1\cup \bar{D}_2, [H_3]_{2}(z,t,\varsigma)$ is analytic for $\varsigma\in D_3\cup D_4$, and continues to $\bar{D}_3\cup \bar{D}_4$.
\item As $\varsigma\rightarrow\infty$, $[H_j]_{1}(z,t,\varsigma)\rightarrow(1,0)^T$, $[H_j]_{2}(z,t,\varsigma)\rightarrow(0,1)^T.$
\end{itemize}
\end{prop}

\begin{prop}
According to Eq.\eqref{2.25}, one find that $w(\varsigma), W(\varsigma)$ can be expressed by
\begin{subequations}
\begin{align}
&w(\varsigma)=\mathrm{I}-\int_{0}^{\infty}e^{-i(\varsigma^2-\frac{1}{2})(\zeta-z) \hat\sigma}(A_1H_3)(\zeta,0,\varsigma)d\zeta,\label{2.29a}\\
&W^{-1}(\varsigma)=\mathrm{I}+\int_{0}^{T}e^{2i(\varsigma^2-\frac{1}{2})^2(\tau-t)\hat\sigma}(B_1H_1)
(0,\tau,\varsigma)d\tau.\label{2.29b}
\end{align}
\end{subequations}
\end{prop}
Assume that $w(\varsigma),W(\varsigma)$ admits the following $2\times2$ matrix from
\eqa
w(\varsigma)=\left(\begin{array}{cc}
\overline{u(\bar{\varsigma})} & v(\varsigma)\\
\overline{v(\bar{\varsigma})}& u(\varsigma)
\end{array}\right),
W(\varsigma)=\left(\begin{array}{cc}
\overline{U(\bar{\varsigma})} & V(\varsigma)\\
\overline{V(\bar{\varsigma})}& U(\varsigma)\\
\end{array}\right).
\label{2.30}\eeqa
It follows from Eq.\eqref{2.25} and Eqs.\eqref{2.29a}-\eqref{2.29b} that the following key properties are ture:
\begin{itemize}
  \item $$( v(\varsigma), u(\varsigma))^T=[H_3]_2^{D_1\cup D_2}(0,0,\varsigma),\quad
 ( -e^{2i(\varsigma^2-\frac{1}{2})^2T}V(\varsigma), \overline{U(\bar{\varsigma})})^T
 = [H_1]_2^{D_2\cup D_4}(0,t,\varsigma). $$
 \item $$u(-\varsigma)=u(\varsigma),\,\, v(-\varsigma)=-v(\varsigma),\,\,
         U(-\varsigma)=U(\varsigma),\,\, V(-\varsigma)=-V(\varsigma).$$
 \item $$ For\,\, \varsigma\in \mathrm{R},\quad \mathrm{det} w(\varsigma)=1,\quad for\,\,\varsigma\in \mathrm{C}\,\,(\mathrm{Im}(\varsigma^2-\frac{1}{2})^2=0,\, if\, T=\infty),\,\, \mathrm{det} W(\varsigma)=1.$$
 \item $$ As\,\varsigma\rightarrow\infty, \,u(\varsigma)=1+\mathrm{O}(\varsigma^{-1}),\,\,v(\varsigma)=\mathrm{O}(\varsigma^{-1}),
     \,\,U(k)=1+\mathrm{O}(\varsigma^{-1}),\,\, V(\varsigma)=\mathrm{O}(\varsigma^{-1}).$$
 \end{itemize}

\subsection{The basic Riemann-Hilbert problem}

For the convenience of calculation, one introduce the following symbol description
\eqa  \eta(\varsigma)=-(\varsigma^2-\frac{1}{2})z+2(\varsigma^2-\frac{1}{2})^2t,\,
\beta(\varsigma)=u(\varsigma)\overline{U(\bar{\varsigma})}+v(\varsigma)\overline{V(\bar{\varsigma})},\,
\delta(\varsigma)=\frac{v(\varsigma)}{\overline{u(\bar{\varsigma})}},\, \Delta(\varsigma)=-\frac{\overline{V(\bar{\varsigma})}}{u(\varsigma)\beta(\varsigma)},\label{2.31}\eeqa
and the $E(z,t,\varsigma)$ defined by
\begin{subequations}
\begin{align}
& E_{+}(z,t,\varsigma)=([H_3]_1^{D_1\cup D_2}(z,t,\varsigma),
\frac{[H_2]_2^{D_1}(z,t,\varsigma)}{\overline{\beta(\bar{\varsigma})}}),\varsigma\in D_1,\label{2.32a}\\
& E_{-}(z,t,\varsigma)=([H_3]_1^{D_1\cup D_2}(z,t,\varsigma),
\frac{[H_1]_2^{D_2}(z,t,\varsigma)}{\overline{u(\bar{\varsigma})}}),\varsigma\in D_2,\label{2.32b}\\
& E_{-}(z,t,\varsigma)=(\frac{[H_2]_1^{D_3}(z,t,\varsigma)}{\beta(\varsigma)},
[H_3]_2^{D_3\cup D_4}(z,t,\varsigma)),\varsigma\in D_3,\label{2.32c}\\
& E_{+}(z,t,\varsigma)=(\frac{[H_1]_1^{D_4}(z,t,\varsigma)}{u(\varsigma)},
[H_3]_2^{D_3\cup D_4}(z,t,\varsigma)),\varsigma\in D_4.\label{2.32d}
\end{align}
\end{subequations}
Above definitions indicate that
\beq  \mathrm{det}E(z,t,\varsigma)=1, E(z,t,\varsigma)\rightarrow\mathrm{I}, \varsigma\rightarrow\infty. \label{2.33}\eeq

\begin{thm}
Let that $r(z,t)\in \mathcal{S}$, for $\varsigma\in\bar{D}_j, j=1,\ldots,4$, the function $W(z,t,\varsigma)$ defined by Eq.\eqref{2.32a}- Eq.\eqref{2.32d} possess the following jump relation
\eqa E_{-}(z,t,\varsigma)=E_{+}(z,t,\varsigma)G(z,t,\varsigma), \varsigma\in \bar{D}_j, j=1,\ldots,4,\label{2.34}\eeqa
where
\eqa
G(z,t,\varsigma)=\left\{ \begin{array}{l}
G_1(z,t,\varsigma), \qquad\qquad\qquad   \mathrm{arg}(\varsigma^2-\frac{1}{2})=\frac{\pi}{2},\\
G_2(z,t,\varsigma)=G_3G_4^{-1}G_1, \quad \mathrm{arg}(\varsigma^2-\frac{1}{2})=\pi,\\
G_3(z,t,\varsigma), \qquad\qquad\qquad \mathrm{arg}(\varsigma^2-\frac{1}{2})=\frac{3\pi}{2},\\
G_4(z,t,\varsigma), \qquad\qquad\qquad   \mathrm{arg}(\varsigma^2-\frac{1}{2})=0,
\end{array}\right.\label{2.35}\eeqa
and
\eqa &&
G_1(z,t,\varsigma)=\left(\begin{array}{cc}
  1 & 0 \\
\Delta(\varsigma)e^{2i\eta(\varsigma)} & 1
 \end{array}\right),\nn\\&&
G_3(z,t,\varsigma)=\left(\begin{array}{cc}
1 &\overline{\Delta(\bar{\varsigma})}e^{-2i\eta(\varsigma)}\\
0 & 1
\end{array} \right),\nn\\&&
G_4(z,t,\varsigma)=\left(\begin{array}{cc}
1 & -\delta(\varsigma)e^{-2i\eta(\varsigma)} \\
\overline{\delta(\bar{\varsigma})}e^{2i\eta(\varsigma)} & 1-|\delta(\varsigma)|^2
\end{array}\right).\nn\eeqa
\end{thm}
\textbf{Proof.} Following\cite{Lenells2008}, one can get Eq.\eqref{2.35}.
\begin{ass}
Set that
\begin{itemize}
\item $u(\varsigma)$ enjoy $2a$ possible single roots $\{\xi_j\}_{j=1}^{2a}$, $2a=2a_1+2a_2$,
  if one let $\{\xi_j\}_1^{2a_1}\in D_4,$ then $\{\bar{\xi}_j\}_1^{2a_2}\in D_2.$
\item $\beta(\varsigma)$ enjoy $2b$ possible single roots $\{\mu_j\}_{j=1}^{2b}$, $2b=2b_1+2b_2$,
  if one let $\{\mu_j\}_1^{2b_1}\in D_1,$ then $\{\bar{\mu}_j\}_1^{2b_2}\in D_3.$
\item The intersection for above possible single roots of $\beta(\varsigma)$ and $u(\varsigma)$ is empty.
\end{itemize}
\end{ass}

\begin{prop}
(The residue conditions) If one remember that $\dot{\beta}(\varsigma)=\frac{d\beta}{d\varsigma}$, the function $W(z,t,\varsigma)$ defined by Eq.\eqref{2.32a}- Eq.\eqref{2.32d} also possess residue conditions as follows
\begin{subequations}
\begin{align}
& \mathrm{Res} \{[E(z,t,\varsigma)]_{1} , \xi_j\}
=\frac{1}{v(\xi_j)\dot{u}(\xi_j)}e^{2i\eta(\xi_j)}[E(z,t,\xi_j)]_{2}, j=1,\cdots,2a_1.\label{2.40a}\\
& \mathrm{Res} \{[E(z,t,\varsigma)]_{2} ,  \bar{\xi}_j\}
=-\frac{1}{\overline{v(\xi_j)}\overline{\dot{u}(\xi_j)}}e^{-2i\eta(\bar{\xi}_j)}[E(z,t,\bar{\xi}_j)]_{1}, j=1,\cdots,2a_2.\label{2.40b}\\
& \mathrm{Res} \{[E(z,t,\varsigma)]_{1} , \mu_j \}
=-\frac{\overline{V(\bar{\mu}_j)}}{u(\mu_j)\dot{\beta}(\mu_j)}e^{2i\eta(\mu_j)}[E(z,t,\mu_j)]_{1}, j=1,\cdots,2b_1.\label{2.40c}\\
& \mathrm{Res} \{[E(z,t,\varsigma)]_{2} , \ \bar{\mu}_j\}=\frac{V(\bar{\mu}_j)}{\overline{u(\mu_j)}\overline{\dot{\beta}(\mu_j)}}e^{-2i\eta(\bar{\mu}_j)}[E(z,t,\bar{\mu_j})]_{2}, j=1,\cdots,2b_2.\label{2.40d}
\end{align}
\end{subequations}
\end{prop}
\textbf{Proof.} Following\cite{Lenells2008}, we can get above residue conditions.

\subsection{ The inverse problem}

The inverse problem is mainly to reconstruct the potential function $r(z,t)$ from eigenfunctions $\{H_j(z,t,\varsigma)\}_1^3$. From section 2.1, we are not difficult to find
$Q_1^{(od)}=-\frac{i}{2}RQ_0\sigma$,
and Eq.\eqref{2.9} is the solution to Eq.\eqref{2.7}, which means
\eqa r(z,t)=2ih(z,t)e^{2i\int_{(0,0)}^{(z,t)}\Theta},\label{2.41}\eeqa
where $H(z,t,\varsigma)$ related to $\Phi(z,t,\varsigma)$ by Eq.\eqref{2.15}, which is defined as
\eqa H(z,t,\varsigma)=\mathrm{I}+\frac{h^{(1)}(z,t)}{\varsigma}+\frac{h^{(2)}(z,t)}{\varsigma^2}+O(\frac{1}{\varsigma^3}),\,
\varsigma\rightarrow\infty,\label{2.41a}\eeqa
if one write $h(z,t)$ for $h^{(1)}_{12}(z,t)$, above Eq\eqref{2.41a} is the solution of Eq.\eqref{2.16}. It follows from Eq.\eqref{2.41} and its complex conjugate that
\eqa r\overline{r}=4|h|^2,\quad \overline{r}r_z-r\overline{r}_z=4(\overline{h}h_z-h\overline{h}_z)-16i|h|^4.\nn\eeqa
Then, the 1-form $\Theta$ is given by Eq.\eqref{2.14} can be express by $g(z,t)$
\eqa
\Theta=|h|^2dz-(2|h|^4+i(\overline{h}h_z-h\overline{h}_z))dt.
\label{2.42}\eeqa
Hence, one can solve the inverse problem following steps:
\begin{description}
  \item[(i)] First of all, using any one of the functions $\{H_j(z,t,\varsigma)\}_1^3$ to calculate $h(z,t)$ according to following equation
\eqa h(z,t)=\lim_{\varsigma\rightarrow\infty}(\varsigma H_j(z,t,\varsigma))_{12}.\nn\eeqa
  \item[(ii)] Secondly, get the 1-form $\Theta(z,t)$ from Eq.\eqref{2.42}.
  \item[(iii)] At last, calculate the potential function $r(z,t)$ according to Eq.\eqref{2.41}.
\end{description}

\subsection{The global relation}

In this subsection, we give the spectral functions $u(\varsigma),v(\varsigma), U(\varsigma), V(\varsigma)$ are not independent but admits a important relation. In fact, the integral of the 1-form $J(z,t,\varsigma)$ is defined by the Eq.\eqref{2.17a} is vanished for the boundary of the region ${(\zeta,\tau): 0<\zeta<\infty, 0<\tau<t}$. If one let $H(z,t,\varsigma)=H_3(z,t,\varsigma)$ in the 1-form, one obtain
\eqa && \int_{\infty}^{0}e^{-i(\varsigma^2-\frac{1}{2})\zeta\hat\sigma}(A_1H_3)(\zeta,0,\varsigma)d\zeta
+\int_{0}^{t}e^{2i(\varsigma^2-\frac{1}{2})^2\tau\hat\sigma}(B_1H_3)(0,\tau,\varsigma)d\tau
+e^{2i(\varsigma^2-\frac{1}{2})^2 t\hat\sigma}
\int_{0}^{\infty}e^{-i(\varsigma^2-\frac{1}{2})\zeta\hat\sigma}(A_1H_3)(\zeta,t,\varsigma)d\zeta\nn\\&&
=\lim_{z\rightarrow\infty}e^{-i(\varsigma^2-\frac{1}{2})z\hat\sigma}
\int_{0}^{t}e^{2i(\varsigma^2-\frac{1}{2})^2\tau\hat\sigma}(B_1H_3)(z,\tau,\varsigma)d\tau.
\label{2.44}\eeqa
On the one hand, as result of $w(\varsigma)=H_3(0,0,\varsigma)$, together with Eq.\eqref{2.28d}, one can see that the first term of the Eq.\eqref{2.44} is
$$w(\varsigma)-I.$$
Set $z=0$ in the Eq.\eqref{2.24a}, one get
\eqa H_3(0,\tau,\varsigma)=H_1(0,\tau,\varsigma)e^{-2i(\varsigma^2-\frac{1}{2})^2\tau\hat\sigma}w(\varsigma),  \label{2.45}\eeqa
then
\eqa e^{2i(\varsigma^2-\frac{1}{2})^2\tau\hat\sigma}(B_1H_3)(0,\tau,\varsigma)
=[e^{2i(\varsigma^2-\frac{1}{2})^2\tau\hat\sigma}(B_1H_1)(0,\tau,\varsigma)]w(\varsigma).  \label{2.46}\eeqa
On the other hand, it follows from Eq.\eqref{2.46} and Eq.\eqref{2.28a} that the second term of the Eq.\eqref{2.44} is
\eqa \int_{0}^{t}e^{2i(\varsigma^2-\frac{1}{2})^2\tau\hat\sigma}(B_1H_3)(0,\tau,\varsigma)d\tau
=[e^{2i(\varsigma^2-\frac{1}{2})^2t\hat\sigma}B_1H_1(0,t,\varsigma)-I]w(\varsigma).  \nn\eeqa
Letting $r(z,t)\in \mathcal{S}$ for $z\rightarrow\infty$, then, Eq.\eqref{2.44} turn into
\eqa W^{-1}(t,\varsigma)w(\varsigma)+e^{2i(\varsigma^2-\frac{1}{2})^2t\hat\sigma}\times
\int_{0}^{\infty}e^{-i(\varsigma^2-\frac{1}{2})\zeta\hat\sigma}(A_1H_3)(\zeta,t,\varsigma)d\zeta=I,\label{2.48}\eeqa
where the first column of Eq.\eqref{2.48} is valid for $\varsigma^2-\frac{1}{2}$ in the lower half-plane and the second column of Eq.\eqref{2.48} is valid for $\varsigma^2-\frac{1}{2}$ in the upper half-plane, and $W(t,\varsigma)$ is given by
\eqa W^{-1}(t,\varsigma)=e^{2i(\varsigma^2-\frac{1}{2})^2t\hat\sigma}H_1(0,t,\varsigma),\nn\eeqa
if one let $t=T$ and remember that $W(\varsigma)=W(T,\varsigma)$, one find that the Eq.\eqref{2.48} equal to
\eqa W^{-1}(\varsigma)w(\varsigma)+e^{2i(\varsigma^2-\frac{1}{2})^2T\hat\sigma}\times
\int_{0}^{\infty}e^{-i(\varsigma^2-\frac{1}{2})\zeta\hat\sigma}(A_1H_3)(\zeta,T,\varsigma)d\zeta=I.\label{2.49}\eeqa
Hence, the (12)-component of Eq.\eqref{2.49} is the so-called global relation, which is given by
\eqa u(\varsigma)V(\varsigma)-U(\varsigma)v(\varsigma)=e^{4i(\varsigma^2-\frac{1}{2})^2T}c^{+}(\varsigma), \mathrm{Im} \varsigma^2\geq0,\label{2.50}\eeqa
where $c^{+}(\varsigma)$ defined by
\eqa c^{+}(\varsigma)=\int_{0}^{\infty}e^{-2i(\varsigma^2-\frac{1}{2})\zeta}(A_1H_3)_{12}(\zeta,T,\varsigma)d\zeta.\label{2.51}\eeqa

\section{The spectral functions}

\begin{de}
 ($u(\varsigma)$ and $v(\varsigma)$) We assume that $r_{0}(z)=r(z,0)\in\mathbb{S}$, and defined the map
$$\mathbb{L}_1: \{r_0(z)\}\rightarrow \{u(\varsigma),v(\varsigma) \},$$
by
$$ (v(\varsigma), \, u(\varsigma))^T={[H_3]}_2^{D_3\cup D_4}(z,0,\varsigma), \mathrm{Im}(\varsigma^2-\frac{1}{2})\geq0,$$
where the expression of $H_3(z,0,\varsigma)$ is
$$H_3(z,0,\varsigma)=\mathrm{I}-\int_{z}^{\infty }
e^{-i(\varsigma^2-\frac{1}{2})(\zeta-z)\hat{\sigma}}(A_1H_3)(\zeta,0,\varsigma)d\zeta,$$
with $A_1(z,0,\varsigma)$ is expressed by Eq.\eqref{2.29a}.
\end{de}

\begin{prop}
The $u(\varsigma)$ and $v(\varsigma)$) possess the following properties
\begin{description}
  \item[(i)] $u(\varsigma)$ and $v(\varsigma)$) are analytic and bounded for $\mathrm{Im}(\varsigma^2-\frac{1}{2})>0$ and continue for $\mathrm{Im}(\varsigma^2-\frac{1}{2})\geq0$.
  \item[(ii)]$u(\varsigma)=1+O(\frac{1}{\varsigma}),v(\varsigma)=O(\frac{1}{\varsigma})$ as $\varsigma\rightarrow\infty$, $\mathrm{Im}(\varsigma^2-\frac{1}{2})\geq0$.
  \item[(iii)] $u(\varsigma)\overline{u(\bar{\varsigma})}-v(\varsigma)\overline{v(\bar{\varsigma})}=1$, $\varsigma^2\in \mathbb{R}$.
  \item[(iv)] $u(-\varsigma)=u(\varsigma),v(-\varsigma)=-v(\varsigma)$, $\mathrm{Im}(\varsigma^2-\frac{1}{2})\geq0$.
  \item[(v)] The inverse map of $\mathbb{L}_1$ is $\mathbb{L}_1^{-1}=:\mathbb{P}_1: \{u(\varsigma),v(\varsigma) \}\rightarrow \{r_0(z)\}$, is defined by
$$r_0(z)=2ih(z)e^{2i\int_0^z|h(\zeta)|^2d\zeta},\\
h(z)=\lim_{\varsigma\rightarrow\infty}(\varsigma E^{(z)}(z,\varsigma))_{12},$$
where $E^{(z)}(z,\varsigma)$ meets the following Riemann-Hilbert problem.
\end{description}
\end{prop}

\begin{itemize}
  \item $E^{(z)}(z,\varsigma)=\left\{ \begin{array}{l}
  E_{-}^{(z)} (z,\varsigma),  \mathrm{Im}(\varsigma^2-\frac{1}{2})\leq0, \\
  E_{+}^{(z)} (z,\varsigma),  \mathrm{Im}(\varsigma^2-\frac{1}{2})\geq0,
\end{array}\right.$ is a piecewise analytical function.
  \item $E_{-}^{(z)}(z,\varsigma)=E_{+}^{(z)}(z,\varsigma)G^{(z)}(z,\varsigma)$, $(\varsigma^2-\frac{1}{2})\in\mathbb{R}$, and
\eqa G^{(z)} (z,\varsigma)=\left(\begin{array}{cc}
1 & -\delta(\varsigma)e^{2i(\varsigma^2-\frac{1}{2})z} \\
\overline{\delta(\bar{\varsigma})}e^{-2i(\varsigma^2-\frac{1}{2})z} & 1-|\delta(\varsigma)|^2
\end{array}\right). \label{3.2}\eeqa
  \item $E^{(z)} (z,\varsigma)=\mathrm{I}+O(\frac{1}{\varsigma}), \varsigma\rightarrow\infty.$
  \item $u(k)$ possess $2a$ simple zeros $\{\xi_j\}_{1}^{2a}$, $2a=2a_1+2a_2$, if $\xi_j\in D_4, j=1,2,\cdots,2a_1$, then $\xi_j\in D_3, j=1,2,\cdots,2a_2$.
  \item The first column of $E_{+}^{(z)}(z,\varsigma)$ enjoy simple poles at $\varsigma=\{\bar{\xi}_j\}_{1}^{2a_2}$. The second column of $E_{-}^{(z)}(z,\varsigma)$ enjoy simple poles at $\varsigma=\{\xi_j\}_{1}^{2a_1}$.
The corresponding residues are expressed by
\begin{subequations}
\begin{align}
& \mathrm{Res} \{[E^{(z)}(z,\varsigma)]_{1},\xi_j\}
=\frac{e^{-2i(\xi_j^2-\frac{1}{2})z}}{\dot{u}(\xi_j)v(\xi_j)}[E^{(z)}(z,\xi_j)]_{2}, j=1,2,\cdots,2a_1, \label{3.4a}\\
& \mathrm{Res} \{[E^{(z)}(z,\varsigma)]_{2} ,  \bar{\xi}_j \}
=\frac{e^{2i(\xi_j^2-\frac{1}{2})z}}{\overline{\dot{u}(\xi_j)}\overline{v(\xi_j)}}
[E^{(z)}(z,\bar{\xi}_j)]_{1}, j=1,2,\cdots,2a_2.\label{3.4b}
\end{align}
\end{subequations}
\end{itemize}

\begin{de}
( $U(\varsigma)$ and $V(\varsigma)$). Similarly, we also set that $s_{0}(t),s_{1}(t)\in \mathbb{S}$, and define the map
$$\mathbb{L}_2: \{s_0(t),s_1(t)\}\rightarrow \{U(\varsigma),V(\varsigma) \},$$
by
$$(V(\varsigma),U(\varsigma))^T={[H_1]}_2^{D_4}(z,0,\varsigma),$$
where the expression of $H_1(0,t,\varsigma)$ is
$$H_1(0,t,\varsigma)=\mathrm{I}-\int_{t}^{T}e^{2i(\varsigma^2-\frac{1}{2})^2(\tau-t)\hat{\sigma}}(B_1H_1)(0,\tau,\varsigma)d\tau,$$
and $B_1(0,t,\varsigma)$ is expressed by Eq.\eqref{2.29b}.
\end{de}

\begin{prop}
The $U(\varsigma)$ and $V(\varsigma)$ admits the properties as follows
\begin{description}
  \item[(i)] $U(\varsigma)$ and $V(\varsigma)$ are analytic and bounded for $\mathrm{Im}(\varsigma^2-\frac{1}{2})^2\geq0$, if $T=\infty$, $U(\varsigma)$ and $V(\varsigma)$ ar defined only for $\mathrm{Im}(\varsigma^2-\frac{1}{2})^2\geq0$.
  \item[(ii)]$U(\varsigma)=1+O(\frac{1}{\varsigma}),V(\varsigma)=O(\frac{1}{\varsigma})$ as $\varsigma\rightarrow\infty$, $\mathrm{Im}(\varsigma^2-\frac{1}{2})^2\geq0$.
  \item[(iii)] $U(\varsigma)\overline{U(\bar{\varsigma})}-V(\varsigma)\overline{V(\bar{\varsigma})}=1$, $\varsigma\in \mathbb{C}((\varsigma^2-\frac{1}{2})^2\in \mathbb{R},\,\,if\,\,T=\infty)$.
  \item[(iv)] $U(-\varsigma)=U(\varsigma),V(-\varsigma)=-V(\varsigma)$.
  \item[(v)] The inverse map of $\mathbb{L}_2$ is $\mathbb{L}_2^{-1}=\mathbb{P}_2: \{U(\varsigma),V(\varsigma) \}\rightarrow \{s_0(t),s_1(t)\}$, is defined by
\eqa &&
s_0(t)=2ih^{(1)}_{12}(t)e^{2i\int_{0}^{t}\Theta_2(\tau)d\tau},\nn\\&&
s_1(t)=(4h_{12}^{(3)}(t)-|s_0(t)|^2h_{12}^{(1)}(t))e^{2i\int_{0}^{t}\Theta_2(\tau)d\tau}
-2is_0(t)h_{22}^{(2)}(t)-is_0(t)+\frac{1}{2}is_0(t)|s_0(t)|^2, \label{3.6}\eeqa
where
$$\Theta_2(\tau)=6|h_{12}^{(1)}|^4-2(\bar{h}_{12}^{(1)}h_{12}^{(3)}+h_{12}^{(1)}\bar{h}_{12}^{(3)})
-2|h_{12}^{(1)}|^2\mathrm{Re}[h_{22}^{(2)}],$$
and the function $h^{(j)}(t),j=1,2,3$ admits the following asymptotic expansion
$$E^{(t)}(t,\varsigma)=\mathrm{I}+\frac{h^{(1)}(t)}{\varsigma}+\frac{h^{(2)}(t)}{\varsigma^2}+\frac{h^{(3)}(t)}{\varsigma^3}
+O(\frac{1}{\varsigma^4}),\,\varsigma\rightarrow\infty,$$
where $E^{(t)}(t,\varsigma)$ meets the following Riemann-Hilbert problem.
\end{description}
\end{prop}
\begin{itemize}
 \item $E^{(t)}(t,\varsigma)=\left\{
\begin{array}{ll}
  E_{-}^{(t)} (t,\varsigma), & \mathrm{Im}(\varsigma^2-\frac{1}{2})^2\leq0, \\
  E_{+}^{(t)} (t,\varsigma), & \mathrm{Im}(\varsigma^2-\frac{1}{2})^2\geq0,
\end{array}
\right.$ is a piecewise analytical function.
\item $E_{-}^{(t)} (t,\varsigma)=E_{+}^{(t)} (t,\varsigma)G^{(t)} (t,\varsigma)$, $(\varsigma^2-\frac{1}{2})^2\in \mathbb{R}$, and
\eqa
G^{(t)} (t,\varsigma)=\left(\begin{array}{cc}
1 & -\frac{V(\varsigma)}{\overline{U(\overline{\varsigma})}}e^{-4i(\varsigma^2-\frac{1}{2})^2t} \\
\frac{\overline{V(\overline{\varsigma})}}{U(\varsigma)}e^{4i(\varsigma^2-\frac{1}{2})^2t} & \frac{1}{U(\varsigma)\overline{U(\overline{\varsigma})}}
\end{array}\right).
\label{3.8}\eeqa
  \item $E^{(t)} (T,\varsigma)=\mathrm{I}+O(\frac{1}{\varsigma}), \varsigma\rightarrow\infty.$
  \item $U(\varsigma)$ possess $2k$ simple zeros $\{\varepsilon_j\}_{1}^{2k}$, $2k=2k_1+2k_2$
  if $\mathrm{Im}(\varepsilon_j^2-\frac{1}{2})^2>0, j=1,2,\cdots,2k_1$,
  then $\mathrm{Im}(\varepsilon_j^2-\frac{1}{2})^2<0, j=1,2,\cdots,2k_2$.
  \item The first column of $E_{+}^{(t)}(t,\varsigma)$ enjoy simple poles at $\varsigma=\{\bar{\varepsilon}_j\}_{1}^{2k_2}$, the second column of $E_{-}^{(t)}(t,\varsigma)$ enjoy simple poles at $\varsigma=\{\varepsilon_j\}_{1}^{2k_2}$.
The corresponding residues are expressed by
\begin{subequations}
\begin{align}
& \mathrm{Res} \{[E^{(t)}(t,\varsigma)]_{1} , \varepsilon_j \}
=\frac{e^{4i(\varepsilon_j^2-\frac{1}{2})^2t}}{\dot{U}(\varepsilon_j)V(\varepsilon_j)}
[E^{(t)}(t,\varepsilon_j)]_{2}, j=1,2,\cdots,2k_1,\label{3.5a}\\
& \mathrm{Res} \{[E^{(t)}(t,k)]_{2} ,  \bar{\varepsilon}_j \}
=\frac{e^{-4i(\bar{\varepsilon}_j^2-\frac{1}{2})^2t}}
{\overline{\dot{U}(\bar{\varepsilon}_j)}\overline{V(\bar{\varepsilon}_j)}}
[E^{(t)}(t,\bar{\varepsilon}_j)]_{1}, j=1,2,\cdots,2k_2.\label{3.5b}
\end{align}
\end{subequations}
\end{itemize}

\section{The Riemann-Hilbert problem}

\begin{thm}
Set that $r_0(z)\in \mathcal{S}(\mathbb{R^{+}})$, the functions $w(\varsigma)$ and $W(\varsigma)$ are defined by $u(\varsigma)$, $v(\varsigma)$, $U(\varsigma), V(\varsigma)$ are given by Eq.\eqref{2.30}, respectively. Assume that functions $u(\varsigma)$ and $\beta(\varsigma)$ possible simple zeros are showed in \textbf{Assumption 2.4}. Therefore, the function $E(z,t,\varsigma)$ conform to the following Riemann-Hilbert problem:
\begin{itemize}
\item $E(z,t,\varsigma)$ is the slice analytic function for $\varsigma\in D_j$ and continues to $\bar{D}_j,(j=1,\ldots,4)$.
\item $E(z,t,\varsigma)$ come into being jumps on the curves $\{\bar{D}_j\}_1^4$ and admits the jump relation given by \textbf{Theorem 2.3}.
\item $E(z,t,\varsigma)=\mathrm{I}+\mathrm{O}(\frac{1}{\varsigma}),\,\varsigma\rightarrow\infty$.
\item $E(z,t,\varsigma)$ meets the residue conditions given by \textbf{Proposition 2.5}.
\end{itemize}
Hence, the function $E(z,t,\varsigma)$ is only existing. Then, one can using $E(z,t,\varsigma)$ to define $r(z,t)$ as follows
\eqa &&r(z,t)=2ih(z,t)e^{2i\int_{(0,0)}^{(z,t)}\Theta},\quad
h(z,t)=\lim_{\varsigma\rightarrow\infty}(\varsigma E(z,t,\varsigma))_{12},\nn\\&&
\Theta=|h|^2dz-(2|h|^4+i(\overline{h}h_z-h\overline{h}_z))dt\label{4.1} \eeqa
thus, the function $r(z,t)$ is a solution of the CLL-NLS equation\eqref{1.2}. Furthermore, $r(z,0)=r_0(z),\,\,r(0,t)=s_0(t),\,\,r_z(0,t)=s_1(t).$
\end{thm}

\textbf{Proof.} Indeed, one can demonstrate that above Riemann-Hilbert problem following \cite{Lenells2008}.

\section{Conclusions and discussions}

In this paper, one use UT approach to discuss the IBVPs of the CLL-NLS equation \eqref{1.2}, one can also discuss Eq.\eqref{1.2} on a finite interval, and with help of the Deift-Zhou method \cite{Deift1993} to analyze the asymptotic behavior for Eq.\eqref{1.2}. Since the RH problem is equivalent to Gel'fand-Levitan-Marchenko(GLM) theory, one can obtain the soliton solution of Eq.\eqref{1.2} by solving the GLM equation following\cite{ZYS2020}, which are our future investigation work.

\section*{Appendix:  Recovering $s_0(t)$ and $s_1 (t)$}

\begin{appendices}
In this appendix, one will derive $s_0(t)$ and $s_1 (t)$ from $E^{(t)}$ lead to Eq.\eqref{3.6}. Assume that $H(z,t,\varsigma)$ is a solution of Eq.\eqref{2.16}. Substituting Eq.\eqref{2.9} into Eq.\eqref{2.6b} and comparing the coefficient for $O(\varsigma)$ yield
\eqa i\sigma R_zQ_0=-4iQ_3^{(od)}\sigma+2RQ_2^{(d)}+iR^2\sigma Q_1^{(od)}+RQ_0+\frac{1}{2}R^3Q_0. \label{A1} \eeqa
where $\Phi(z,t,\varsigma)$ is the solution of Eq.\eqref{2.7} enjoy following form
\beq \Phi(z,t,\varsigma)=Q_0+\frac{Q_1}{\varsigma}+\frac{Q_2}{\varsigma^2}+\frac{Q_3}{\varsigma^3}+O(\frac{1}{\varsigma^4}), \varsigma\rightarrow\infty.\nn\eeq
Since $\Phi(z,t,\varsigma)$ is related to $H(z,t,\varsigma)$ is defined by Eq.\eqref{2.15}, we have written
\eqa
Q_0=\left(\begin{array}{cc}
Q_{11}^{(0)}&0\\
0 & Q_{22}^{(0)}
\end{array}\right),
H(z,t,\varsigma)=\left(\begin{array}{cc}
H_{11} & H_{12}\\
H_{21} & H_{22}
\end{array}\right), \nn\eeqa
then, we have
\eqa
\Phi(z,t,\varsigma)=\left(\begin{array}{cc}
Q_{11}^{(0)}H_{11} & Q_{22}^{(0)}e^{2i\int_{(0,0)}^{(z,t)}\Theta}H_{12}\\
Q_{11}^{(0)}e^{-2i\int_{(0,0)}^{(z,t)}\Theta}H_{21} & Q_{22}^{(0)}H_{22}
\end{array}\right). \nn\eeqa
If one set
\beq H(z,t,\varsigma)=I+\frac{h^{(1)}}{\varsigma}+\frac{h^{(2)}}{\varsigma^2}+\frac{h^{(3)}}{\varsigma^3}+O(\frac{1}{\varsigma^4}), \varsigma\rightarrow\infty.\nn\eeq
then the (12)-entry of Eq.\eqref{A1} gives
\eqa r_z=4h_{12}^{(3)}e^{2i\int_{(0,0)}^{(z,t)}\Theta}-2irh_{22}^{(2)}
-|r|^2h_{12}^{(1)}e^{2i\int_{(0,0)}^{(z,t)}\Theta}-ir+\frac{1}{2}ir|r|^2
\label{A2}\eeqa
Take the complex conjugate yield
\eqa \bar{r}_z=4\bar{h}_{12}^{(3)}e^{-2i\int_{(0,0)}^{(z,t)}\Theta}+2i\bar{r}\bar{h}_{22}^{(2)}
-|r|^2\bar{h}_{12}^{(1)}e^{-2i\int_{(0,0)}^{(z,t)}\Theta}+i\bar{r}-\frac{1}{2}i\bar{r}|r|^2 \label{A3}\eeqa
At the same time, from Eq.\eqref{2.41} one finds
\eqa r(z,t)=2ih_{12}^{(1)}e^{2i\int_{(0,0)}^{(z,t)}\Theta},\,\,
\bar{r}(z,t)=-2i\bar{h}_{12}^{(1)}e^{-2i\int_{(0,0)}^{(z,t)}\Theta},
 \label{A4}\eeqa
It follows from Eqs.\eqref{A2}-\eqref{A4} that
\eqa \bar{r}r_z-r\bar{r}_z=-8i(\bar{h}_{12}^{(1)}h_{12}^{(3)}+h_{12}^{(1)}\bar{h}_{12}^{(3)})
+4ir\bar{r}h_{12}^{(1)}\bar{h}_{12}^{(1)}-2ir\bar{r}\mathrm{Re}[h_{22}^{(2)}]-2ir\bar{r}+ir^2\bar{r}^2, \label{A5}\eeqa
which means that the coefficient $\Theta_2=\frac{1}{8}|r|^4-\frac{i}{4}(\bar{r}r_z-r\bar{r}_z)$ of $\mathrm{d}t$ in the differential form $\Theta$ is defined in Eq.\eqref{2.14a} can be expressed as
\eqa \Theta_2=\frac{1}{8}|r|^4-2(\bar{h}_{12}^{(1)}h_{12}^{(3)}+h_{12}^{(1)}\bar{h}_{12}^{(3)})
+r\bar{r}h_{12}^{(1)}\bar{h}_{12}^{(1)}-\frac{1}{2}r\bar{r}\mathrm{Re}[h_{22}^{(2)}],
\label{A6}\eeqa
Owing to $r\bar{r}=4|h_{12}^{(1)}|^2,$ we calculate Eq.\eqref{A2},\eqref{A4}-\eqref{A7} at $z=0$ yields
\eqa && s_0(t)=2ih^{(1)}_{12}(t)e^{2i\int_{0}^{t}\Theta_2(\tau)d\tau}, \nn\\&&
s_1(t)=(4h_{12}^{(3)}(t)-|s_0(t)|^2h_{12}^{(1)}(t))e^{2i\int_{0}^{t}\Theta_2(\tau)d\tau}
-2is_0(t)h_{22}^{(2)}(t)-is_0(t)+\frac{1}{2}is_0(t)|s_0(t)|^2, \label{A7}\eeqa
with
\eqa \Theta_2(\tau)=6|h_{12}^{(1)}|^4-2(\bar{h}_{12}^{(1)}h_{12}^{(3)}+h_{12}^{(1)}\bar{h}_{12}^{(3)})
-2|h_{12}^{(1)}|^2\mathrm{Re}[h_{22}^{(2)}], \label{A8}\eeqa
and the function $h^{(j)}(t),j=1,2,3$ determined by following asymptotic expansion
\eqa E^{(t)}(t,\varsigma)=\mathrm{I}+\frac{h^{(1)}(t)}{\varsigma}+\frac{h^{(2)}(t)}{\varsigma^2}+\frac{h^{(3)}(t)}{\varsigma^3}
+O(\frac{1}{\varsigma^4}),\,\varsigma\rightarrow\infty.\label{A9}\eeqa
\end{appendices}

\section*{Acknowledgements}
This work is supported by the NSFC under Grant Nos. 11601055, 11805114 and 11975145, NSF of Anhui Province under Grant No.1408085QA06, Natural Science Research Projects of Anhui Province under Grant Nos. KJ2019A0637 and gxyq2019096.

\end{document}